\documentclass{article}

\usepackage{amssymb}
\usepackage{amsmath}

\newcommand{\Prob}{\mathrm{Prob}}
\newcommand{\smfrac}[2]{\mbox{$\frac{#1}{#2}$}}

\newcommand{\F}{\mathbb{F}}
\newcommand{\Had}{\texttt{H}}
\renewcommand{\H}{\mathcal{H}}
\newcommand{\N}{\mathbb{N}}
\newcommand{\Z}{\mathbb{Z}}
\newcommand{\ket}[1]{|#1\rangle}
\newcommand{\braket}[2]{\langle#1|#2\rangle}
\newcommand{\inner}[2]{(#1,#2)}
\newcommand{\C}{\mathbb{C}}
\newcommand{\ox}{\otimes}

\newcommand{\OR}{\texttt{OR}}
\newcommand{\W}{\mathsf{W}}
\newcommand{\U}{\mathsf{U}}
\newcommand{\I}{\mathrm{I}}
\newcommand{\J}{\mathrm{J}}

\newtheorem{theorem}{Theorem}
\newtheorem{lemma}{Lemma}
\newtheorem{definition}{Definition}
\newtheorem{corollary}{Corollary}
\newtheorem{example}{Example}
\newtheorem{fact}{Fact}

\def\squareforqed{\hbox{\rlap{$\sqcap$}$\sqcup$}}
\def\qed{\ifmmode\squareforqed\else{\unskip\nobreak\hfil
\penalty50\hskip1em\null\nobreak\hfil\squareforqed
\parfillskip=0pt\finalhyphendemerits=0\endgraf}\fi}
\newenvironment{proof}{\noindent \textbf{Proof: }}{\qed}

\mathchardef\ordinarycolon\mathcode`\:
\mathcode`\:=\string"8000
\def\vcentcolon{\mathrel{\mathop\ordinarycolon}}
\begingroup \catcode`\:=\active
  \lowercase{\endgroup
  \let :\vcentcolon
  }

\title{\textbf{Quantum Algorithms for 
Weighing Matrices and Quadratic Residues}}
\author{Wim van Dam\thanks{Computer Science Division, Soda Hall, 
University of California, Berkeley, CA 94720 (USA)}\\
UC Berkeley\\ vandam@cs.berkeley.edu}
\begin{document}

\maketitle

\begin{abstract}
In this article we investigate how we can employ the structure of 
combinatorial objects like Hadamard matrices and
weighing matrices to device new quantum algorithms.
We show how the properties of a weighing matrix can be used 
to construct a problem for which the quantum query complexity is 
significantly lower than the classical one.
It is pointed out that this scheme captures both Bernstein \& 
Vazirani's inner-product protocol, as well as Grover's search algorithm.

In the second part of the article we consider Paley's construction
of Hadamard matrices, which relies on the properties of 
quadratic characters over finite fields.
 We design a query problem that uses the Legendre symbol $\chi$ 
(which indicates if an element of a finite 
field $\F_{q}$ is a quadratic residue or not).
It is shown how for a shifted Legendre function
$f_s(i)=\chi(i+s)$, the unknown $s \in \F_{q}$ can be obtained 
exactly with only two quantum calls to $f_s$. This is in sharp contrast 
with the observation that any classical, probabilistic procedure requires 
more than $\log q + \log (\smfrac{1-\varepsilon}{2})$ queries to solve the same problem.
\end{abstract}

\section{Introduction}
The theory of quantum computation investigates how we can 
use quantum mechanical effects to solve computational 
problems more efficiently than we can by classical means. 
So far, the strongest evidence that there is indeed a real 
and significant difference between quantum and classical 
computation is provided by Peter Shor's polynomial-time 
factoring algorithm\cite{S:afqcdlaf}.
Most other quantum complexity results are expressed in the black-box, 
or oracle, setting of computation with various degrees
of separation between the two models. 
The algorithms of---for example---Deutsch\cite{D:qttctpatuqc}, 
Deutsch \& Jozsa\cite{DJ:rsopbq}, Berthiaume \& Brassard\cite{BB:oqc},
Bernstein \& Vazirani\cite{BV:qct}, Simon\cite{S:otpoqc}, 
Grover\cite{G:afqmafds}, and Buhrman \& van Dam\cite{BD:bqqc} 
define problems for which we have a quantum reduction in the query complexity,
whereas the lower bounds of Jozsa\cite{J:ccofcbqp}, Bennett 
\emph{et al.}\cite{BBBV:sawoqc}, and Beals \emph{et al.}\cite{BBCMW:qlbbp} 
show that there are limits to the 
advantage that quantum computation can give us. 
The general picture that has emerged from these results
is that we can only expect a superpolynomial difference between 
classical and quantum computation if we can use the specific 
structure of the problem that we try to solve. 
The promise on the function of Simon's problem 
is a typical example of such a structure that establishes an exponential
quantum improvement over the classical complexity.\cite{S:otpoqc}
It was this same improvement that inspired Shor for his result.

In this article we introduce a general family of structured problems
for which we prove a better-than-classical query complexity.
These results rely heavily on the properties of weighing matrices
(as defined in combinatorics) and encompass the earlier query 
protocols by Grover\cite{G:afqmafds} and 
Bernstein \& Vazirani\cite{BV:qct}.
Following Paley's construction of Hadamard matrices, we also define 
a more specific problem that concerns the determination of 
a shifted Legendre sequence over finite fields.

In the next section we start with a brief overview of the
essential ingredients of quantum computation and some of
the relevant complexity results for the black-box model. 
Section~\ref{sec:hwmatrix} then explains how the theory
of weighing matrices can be used as a source for non-trivial, 
yet structured, unitary operations. 
In the last part of the article (Section~\ref{sec:SLSP}) our attention 
will focus on Raymond Paley's construction of Hadamard matrices 
and the theory of quadratic residues for finite fields that it uses.
This will lead to the definition of a query problem which is
akin to the inner-product problem of Bernstein \& Vazirani\cite{BV:qct}.

\section{Quantum Computation}
We assume the reader to be familiar with the theory of 
quantum computation. (Otherwise, see the standard references
by Berthiaume\cite{B:qc}, Nielsen and Chuang\cite{NC:qcaqi}, 
or Preskill\cite{P:qc}.)
Here we will mainly fix the terminology and notation for the
 rest of the article.

\subsection{Quantum Information Processing}
A system $\psi$ of $n$ quantum bits (qubits) is a superposition
of all  possible $n$-bit strings.  It can therefore be 
represented as a normalized vector (or ``ket'') 
$\ket{\psi}$ in a $2^n$-dimensional Hilbert space:
\begin{eqnarray*}
\ket{\psi}  &=& \sum_{x \in \{0,1\}^n}{\alpha_x\ket{x}},
\end{eqnarray*}
with $\alpha_x \in \C$ and the normalization restriction 
$\sum_x {|\alpha_x|^2}=1$.
The probability of observing the outcome ``$x$'' when measuring
the state $\psi$ equals $|\alpha_x|^2$. 
More general, when we try to determine if $\psi$
 equals the measurement vector 
$\ket{m} = \sum_x {\beta_x\ket{x}}$, we will get an affirmative
answer with probability
\begin{eqnarray*}
\Prob(m|\psi)  \quad :=\quad  |\braket{m}{\psi}|^2  & = &
\left|{\sum_{x\in\{0,1\}^n} {\bar{\beta}_x \alpha_x}}\right|^2
\end{eqnarray*}
(with $\bar{\beta}$ the complex conjugate of $\beta$).
An \emph{orthogonal measurement basis} for an 
$N$-dimensional Hilbert space $\H_N$
is a set $\{m_1,m_2,\dots,m_N\}$ of mutually
orthogonal state vectors $\ket{m_i}$. 
For such a basis it holds that 
$\sum_{i=1}^N{\Prob(m_i|\psi)}=1$,
for every state $\ket{\psi} \in \H_N$, and that if 
$\psi=m_s$ for a certain $s$, then 
$\Prob(m_i|\psi)  = 1$ if $i=s$, and 
$\Prob(m_i|\psi)  = 0$ otherwise. 

The quantum mechanical time evolution of a system $\psi$ 
is a linear transformation that preserves the normalization restriction.
Hence, for a finite-dimensional state space $\H_N$, 
such a transformation can be represented by a unitary matrix 
$M \in \U(N)$, for which we can write
$M\ket{\psi} = \sum_{x=1}^N{\alpha_x M\ket{x}}$.
An example of a one-qubit transformation is the `Hadamard transform', 
which is represented by the unitary matrix
\begin{eqnarray*}
\Had & := & \smfrac{1}{\sqrt{2}}
\left({\begin{array}{cc}
+1 & +1 \\
+1 & -1\\
\end{array}}\right).
\end{eqnarray*}
On the standard zero/one basis for a bit this transformation has the
following effect:
$\Had\ket{0} = \smfrac{1}{\sqrt{2}}(\ket{0}+\ket{1})$ and 
$\Had\ket{1} = \smfrac{1}{\sqrt{2}}(\ket{0}-\ket{1})$.

\subsection{Quantum versus Classical Query Complexity}
Consider a problem that is defined in terms of 
$n$ (unknown) values $f(1),\dots,f(n)$. 
The \emph{(probabilistic) query complexity} of such a problem is the minimum 
number of times that an algorithm has to `consult' the string
$f(1),\dots,f(n)$ to solve the problem (with high probability).
A typical example of this setting is the calculation
of the $\OR$ of $n$ bit values: the question
whether there is an index $i$ with $f(i)=1$.
The classical probabilistic, query complexity of this task 
is $\Omega(n)$, 
whereas in the quantum setting we only need $O(\sqrt{n})$calls to $f$ 
to solve the problem with high probability.  
We therefore say that we have a 
`quadratic' separation between the classical and the
quantum query complexity of the $\OR$ function.
The question which tasks allow a quantum reduction 
in the query complexity (and if so, how much) is a central one 
in quantum complexity research.

\subsection{Some Earlier Results in Quantum Computing}
In this article we are especially concerned 
with the query complexity of procedures that 
prepare a state that depends on the values of 
a black-box function.
For example, how often do we have to read out the
bit values of a function $f:\N\rightarrow\C$ 
if we want to create the state $\sum_i {{f(i)}\ket{i}}$?   
The following lemma shows us that if the range of the function 
is limited to $\{-1,+1\}$, this can be done with a single query. 
\begin{fact}[Phase-kick-back trick]\label{fc:kickback}
Given a function $f:\N\rightarrow\{-1,+1\}$, 
the phase changing transition
$\sum_{i}{\alpha_i\ket{i}} \rightarrow 
\sum_{i}{{f(i)}\alpha_i\ket{i}}$
can be established with only one call to the unknown binary values of $f$. 
\end{fact}
\begin{proof}
(See \cite{CEMM:qar} for the original proof.)
First, attach to the superposition of $\sum_i{\alpha_i\ket{i}}$
the (two qubit state)
\begin{eqnarray*}
\ket{\varphi} & := & 
\smfrac{1}{2}(\ket{0}+\sqrt{-1}\ket{1}-\ket{2}-\sqrt{-1}\ket{3}).
\end{eqnarray*}
Then, in superposition, add modulo $4$ the values $f(i)$
to this register (step $a$). 
Finally, apply a general phase change 
$\ket{\Psi}\rightarrow\sqrt{-1}\ket{\Psi}$ (step $b$).  
 It is straightforward to see that this yields the desired 
phase change according to the equation:
\begin{eqnarray*}
\ket{i}\ox\ket{\varphi}
& \longrightarrow_a & 
\left\{\begin{array}{rl}
-\sqrt{-1}\ket{i}\ox\ket{\varphi} & \mbox{if $f(i)=+1$}\\
\sqrt{-1}\ket{i}\ox\ket{\varphi} & \mbox{if $f(i)=-1$}\\
\end{array}\right. \\
& \longrightarrow_b & 
f(i)\ket{i}\ox\ket{\varphi},
\end{eqnarray*}
for every $i$ in the superposition.
\end{proof}
The usefulness of such a phase-changing operation is made 
clear by the following result, which is mentioned because the
Theorems \ref{thm:qupperb} and \ref{thm:qslsp} of this article 
are of a similar fashion.
In 1993 Bernstein \& Vazirani gave the following
example of a family of functions $g_1,g_2,\dots$ that are 
more easily distinguished with quantum queries to $g$ 
than with classical ones.

\begin{fact}[Inner-Product Problem]
Let the  black-box function 
$g_s: \{0,1\}^n \rightarrow \{0,1\}$ be defined by
$g_s(x) = \inner{x}{s}  :=  \sum_{i=1}^{n}{ s_ix_i \bmod{2}}$,
where $s=(s_1,\dots,s_n) \in \{0,1\}^n$ 
is an unknown $n$-bit mask.
A quantum computer can determine the value $s$ exactly 
with one call to the function $g_s$, whereas any probabilistic, 
classical algorithm needs at least $n+\log(1-\varepsilon)$ queries 
to $g_s$ to perform the same task with an error rate of 
at most $\varepsilon$. 
\end{fact}
\begin{proof}
See \cite{BV:qct} for the original proof by Bernstein \& Vazirani, 
and \cite{CEMM:qar} for the single query version of it.
\end{proof}
The above result uses the unitarity of $\Had^{\ox n}$ and
its connection with the inner-product function.
In Section~\ref{sec:SLSP} of this article we will do a similar thing
for a different family of unitary matrices and the Legendre
 function that it uses.

Another key result in quantum computation is the square-root speed-up
that one can obtain when querying a database for a specific element.
\begin{fact}[Grover's search algorithm]\label{fc:grover}
Let the function values $f(1),\dots,f(n)$ form a string of 
$n-k$ zeros and $k$ ones. 
Knowing $k$, but not the specific entries $s$ for which $f(s)=1$, 
the amplitude changing evolution
\begin{eqnarray*}
\frac{1}{\sqrt{n}}\sum_{j=1}^{n}{\ket{i}}
& \longrightarrow & 
\frac{1}{\sqrt{k}}\sum_{j=1}^{n}{f(i)\ket{i}}
\end{eqnarray*}
can be established exactly with  
$\left\lceil{\smfrac{\pi}{4}\sqrt{\smfrac{n}{k}}}\right\rceil$ 
quantum queries to the function $f$.
\end{fact}
\begin{proof}
See the original article by Lov Grover\cite{G:afqmafds}, 
or better yet, the excellent analysis of it by Boyer 
\emph{et al.}\cite{BBHT:tboqs}
\end{proof}

\section{Hadamard Matrices and Weighing Matrices in Combinatorics}
\label{sec:hwmatrix}
The matrix $\Had$ that we mentioned in the previous section
is---in the context of quantum computation---called the 
`Hadamard matrix'. 
 This terminology is a bit unfortunate because 
the same term has already been used 
in combinatorics to cover a much broader concept.
(See the 1893 article by Jacques Hadamard\cite{H:rduqrad} for the 
origin of this term.)

\begin{definition}[Hadamard matrix in combinatorics]
In combinatorics, a matrix $M \in \{-1,+1\}^{n\times n}$ is called 
a \emph{Hadamard matrix} 
if and only if $M\cdot M^T = n\cdot \I_n$,
where ``$T$'' denotes the transpose of a matrix.
\end{definition}
Obviously, when $M$ is a Hadamard matrix in the above sense, 
then $\smfrac{1}{\sqrt{n}}M$ is a unitary matrix $\in \U(n)$.
Also, if $M_1$ and $M_2$ are Hadamard matrices, then
their tensor product $M_1\otimes M_2$ is a Hadamard matrix as well.
It is a famous open problem if there exists a Hadamard
matrix for every dimension $4k$.

The $\Had^{\ox n}$ matrices  that we encountered in the section
on quantum computation form only a small subset of all the Hadamard matrices 
that we know in combinatorics.
Instead, the matrices $\sqrt{2^n}\cdot\Had^{\otimes n}$ should perhaps be called 
``Hadamard matrices of the Sylvester kind'' after the author who first discussed this 
specific family of matrices.\cite{S:toiomsssatpitomcwatnrotwatton}

The properties of Hadamard matrices (especially the above mentioned 
$4k$-conjecture) is an intensively studied topic in combinatorics, and 
its complexity is impressive given the simple
definition.\cite{CD:tchocd,HSS:oataa,S:alohm,S:alwohmsdbfatatcaisat,SY:hmsabd} 
In 1933, Raymond Paley proved the existence of two families of Hadamard 
matrices that are very different from Sylvester's $2^n$-construction. 
\begin{fact}[Paley construction I and II]
Construction I: For every prime $p$ with $p = 3\bmod{4}$ and every integer $k$,
there exists a Hadamard matrix of dimension $(p^k+1)\times (p^k+1)$.
Construction II: For every prime $p$ with $p = 1\bmod{4}$
and every integer $k$, there exists a Hadamard matrix of dimension
$(2p^k+2) \times (2p^k+2)$.
\end{fact}
\begin{proof}
See the original article \cite{P:oom}, or any other standard
text on combinatorial objects\cite{CD:tchocd,MS:ttoecc,SY:hmsabd}. 
\end{proof}
For here it sufficient to say that Paley's construction uses
the theory of quadratic residues of finite fields $\F_{p^k}$.
Its properties that are relevant for this article are discussed
in the appendix.

We can extend the notion of Hadamard matrices by allowing three 
possible matrix entries $\{-1,+1,0\}$, while still requiring
the $M\cdot M^T \propto \I_n$ restriction. 
We thus reach the following definition.

\begin{definition}[Weighing matrix \cite{CD:tchocd,S:alwohmsdbfatatcaisat}]
In combinatorics, a matrix $M \in \{-1,0,+1\}^{n\times n}$ 
is called a \emph{weighing matrix} if and only if
$M\cdot M^T = k\cdot \I_n$ for some $0\leq k\leq n$.
We will denote the set of such matrices by $\W(n,k)$.
\end{definition}
Every column and row of a $\W(n,k)$ weighing matrix
has $n-k$ zeros, and $k$ entries ``$+1$'' or ``$-1$''.
Clearly, $\W(n,n)$ are the Hadamard matrices again, 
whereas $\W(n,n-1)$ are also called \emph{conference matrices.}
The identity matrix $\I_n$ is an example of a $\W(n,1)$
matrix. 
If $M_1 \in \W(n_1,k_1)$ 
and $M_2 \in \W(n_2,k_2)$, then their tensor product
$M_1\ox M_2$ is an element of $\W(n_1n_2,k_1k_2)$. 
This implies that for every weighing matrix $M \in \W(n,k)$ 
we have in fact a whole family of matrices 
$M^{\ox t} \in \W(n^t,k^t)$, indexed by $t \in\N$.

\begin{example}\label{ex:w43} 
\begin{equation*}
{\left({\begin{array}{rrrr}
+1 & +1 & +1 &  0 \\
+1 & -1 &  0 & +1 \\
+1 &  0 & -1 & -1 \\
 0 & +1 & -1 & +1 
\end{array}}\right)}^{\ox t}
\mbox{ is a $\W(4^t,3^t)$ weighing matrix
for every $t\in\N$.} 
\end{equation*}
\end{example}

The observation that for every $M \in \W(n,k)$ the
matrix $\smfrac{1}{\sqrt{k}}\cdot M \in \U(n)$ is a unitary
matrix makes the connection between combinatorics and 
quantum computation that we explore in this article.
In the next section we will see how the mutually orthogonal
basis of such a matrix can be used for a query efficient
quantum algorithm. The classical lower bound for the
same problem is proven using standard, decision tree arguments.
 
\section{Quantum Algorithms for Weighing Matrices}\label{sec:wmp}
In this section we will describe a general weighing-matrix-problem 
and its quantum solution.
But before doing so, we first mention the following state construction
lemma which follows directly from earlier results on Grover's search 
algorithm.
\begin{lemma}[State construction lemma]\label{lm:create}
Let $f:\{1,\dots,n\}\rightarrow \{-1,0,+1\}$ be a black-box function.
If we know that $k$ of the function values are ``$+1$'' or 
``$-1$'', and the remaining $n-k$ entries are ``$0$'', 
then the preparation of the state
\begin{eqnarray*}
\ket{f} & :=  & \frac{1}{\sqrt{k}}\sum_{i=1}^{n}{f(i)\ket{i}},
\end{eqnarray*}
requires no more than 
$\left\lceil{\smfrac{\pi}{4}\sqrt{\smfrac{n}{k}}}\right\rceil+1$ 
quantum evaluations of the black-box function $f$.
When $k=n$, a single query is sufficient.
\end{lemma}
\begin{proof}
First, we use the amplitude amplification process of
Grover's search algorithm\cite{G:afqmafds} described 
in Fact~\ref{fc:grover} to create exactly the state
$\frac{1}{\sqrt{k}}\sum_{i=1,\dots, n}^{f(i)\neq 0}{\ket{i}}$
with no more than 
$\left\lceil{\smfrac{\pi}{4}\sqrt{\smfrac{n}{k}}}\right\rceil$ 
queries to $f$. (See the article by Boyer \emph{et al.\ }\cite{BBHT:tboqs} 
for a derivation of this upper bound. Obviously, no queries are required
if $k=n$.)
After that, following Fact~\ref{fc:kickback}, one additional $f$-call 
is sufficient to insert the proper amplitudes, 
yielding the desired state $\ket{f}$. 
\end{proof}

\subsection{Weighing Matrix Problem and Its Quantum Solution}
We will now define the central problem of this article,
which assumes the existence of a weighing matrix.
\begin{definition}[Weighing Matrix Problem]\label{def:qproblem}
Let $M$ be a $\W(n,k)$ weighing matrix. 
Define a set of $n$ functions 
$f^M_s:\{1,\dots,n\}\rightarrow\{-1,0,+1\}$ for 
every $s \in \{1,\dots,n\}$ by $f^M_s(i) := M_{si}$.
Given a function $f^M_s$ in the form of a black-box, 
we want to determine the parameter $s$.
The (probabilistic) query complexity of the weighing matrix problem
is the minimum number of calls to the function $f$ that is 
necessary to determine the value  $s$ 
(with error probability at most $1-\varepsilon$).
\end{definition}
With the quantum protocol of Lemma~\ref{lm:create} we can solve this 
problem in a straightforward way.
\begin{theorem}[Quantum procedure for the Weighing Matrix 
Problem]\label{thm:qupperb}
For every weighing matrix $M\in \W(n,k)$ with the corresponding
Weighing Matrix Problem of Definition~\ref{def:qproblem}, 
there exists a quantum algorithm that determines
$s$ exactly with at most 
$\left\lceil{\smfrac{\pi}{4}\sqrt{\smfrac{n}{k}}}\right\rceil+1$ 
queries to $f^M_s$. When $n=k$, the problem can be solved with
one query to the function.
\end{theorem}
\begin{proof}
First, prepare the state $\ket{f^M_s} =  
\smfrac{1}{\sqrt{k}}\sum_{i=1}^{n}{f^M_s(i)\ket{i}}$ with 
$\left\lceil{\smfrac{\pi}{4}\sqrt{\smfrac{n}{k}}}\right\rceil+1$ 
queries to the function $f$ (Lemma~\ref{lm:create}).
After that, measure the state in the basis spanned by the vectors
$\ket{f^M_1},\dots,\ket{f^M_n}$. Because $M$ is a weighing
matrix, this basis is orthogonal and hence the outcome of the 
measurement gives us the value $s$ (via the outcome $\ket{f^M_s}$) 
without error. 
\end{proof}

\subsection{Classical Bounds for Weighing Matrix Problems}
For every possible weighing matrix, the above result establishes
a separation between the quantum and the classical
query complexity of the problem, as is shown by the following 
classical lower bound.
\begin{lemma}[Classical lower bounds for the Weighing Matrix 
Problem]\label{lm:clowerb}
Consider the Weighing Matrix Problem of Definition~\ref{def:qproblem} 
for a matrix $M \in \W(n,k)$. Let $d$ be the number of 
queries used by a classical algorithm that recovers $s$ with an 
error probability of at most $\varepsilon$. 
This query complexity $d$ is bounded from below by
the following three inequalities:
\begin{eqnarray*}
d  &\geq&  \log_3 n + \log_3(1-\varepsilon), \\
d  &\geq&  (1-\varepsilon)\smfrac{n}{k}-\smfrac{1}{k}, \\
d &\geq&  \log(\smfrac{n}{n-k+1})+\log(1-\varepsilon).
\end{eqnarray*}
For the case where $k=n$, this last lower bound equals 
$d \geq \log n + \log(1-\varepsilon)$.
\end{lemma}
\begin{proof} 
We will prove these  bounds by considering the decision trees 
that describe the possible classical protocols.
The procedure starts at the root of the tree and this node 
contains the first index $i$ that the protocol queries
to the function $f$.  Depending on the outcome 
$f(i)\in\{-1,0,+1\}$, the protocol follows one of the 
three outgoing edges to a new node $v$, which
contains the next query index $i_v$. 
This routine is repeated until the procedure reaches
one of the leaves of the tree.  
At that point, the protocol guesses which function it
has been querying. 
With this representation, the depth of such a tree reflects 
the number of queries that the protocol uses, while the number 
of leaves (nodes without outgoing edges) indicates how many 
different functions the procedure can distinguish.  

For a probabilistic algorithm with error probability $\varepsilon$,
we need to use decision trees with at least $(1-\varepsilon)n$
leaves.  
Because the number of outgoing edges cannot be bigger than $3$,
a tree with depth $d$ has maximally $3^d$ leaves. 
This proves the first lower bound via 
$3^d \geq (1-\varepsilon)n$.

For the second and third bound we have to analyze the maximum 
size of the decision tree as it depends on the values $k$ and $n$.
We know that for every index $i_v$, there are only $k$ 
different functions with $f(i_v)\neq 0$.
This implies that at every node $v$ the joint number of leaves 
of the two subtrees associated with the outcomes $f(i_v)=-1$
and $+1$ cannot be bigger than $k$. 
Hence, by considering the path (starting from the root) along the 
edges that correspond to the answers $f(i_v)=0$, we see that a decision 
tree with $d$ queries, can distinguish no more than $dk+1$
functions. The second bound is thus obtained by 
the resulting inequality $dk+1\geq (1-\varepsilon)n$.
(The case $k=1$ is the strongest example of this bound.)

In a similar fashion, we can use the observation that there are 
exactly $n-k$ functions with $f(i_v)=0$ for every node $v$.
Now we should consider the binary subtree that is spanned by the edges
that correspond to the answers $f(i_v)=+1$ and $f(i_v)=-1$.
With depth $d$, this subtree has at most $2^d$ leaves and 
$2^d-1$ internal nodes. 
For the complete tree, each such internal node $v$ gives at most 
$n-k$ additional leaves, which are the functions with $f(i_v)=0$.  
In sum, this tells us that the total tree (with depth $d$) has a maximum
 number of leaves of $2^d+(2^d-1)(n-k)$, leading to the third result:
$d\geq \log(\smfrac{n}{n-k+1})+\log(1-\varepsilon)$.
\end{proof}

\subsection{Additional Remarks}
The above bounds simplify significantly when we express them
as functions of big enough $n$, 
giving us the following table:
\begin{center}
\begin{tabular}{|c|c|c|}
\hline $k$ & quantum upper bound & classical lower bound \\\hline\hline
$o(n)$ & ${\smfrac{\pi}{4}\sqrt{\smfrac{n}{k}}} + 2$ & 
$(1-\varepsilon)\smfrac{n}{k} - O(1)$ 
\\\hline
$\Theta(n) $ & $O(1)$ & $\log_3 n + \log_3(1-\varepsilon)$ 
\\\hline
$n$ & $1$ & $\log n + \log(1-\varepsilon)$ \\\hline
\end{tabular}.
\end{center}

Note that the $n$-dimensional identity matrix is a 
$\W(n,1)$ weighing matrix, and that for this $\I_n$ 
the previous theorem and lemma are just a rephrasing 
(with $k=1$) of the results on Grover's search algorithm 
for exactly one matching entry. The algorithm of 
Bernstein \& Vazirani is also captured by the above  
as the case where $k$ has the maximum value $k=n$ (with the 
weighing matrices $(\sqrt{2}\cdot\Had)^{\ox t} \in \W(2^t,2^t)$). 
Hence we can think of those two algorithms as the 
extreme instances of the more general weighing matrix problem. 

As we phrased it, a weighing matrix $M\in \W(n,k)$ 
gives only one specific problem for which there is a 
classical/quantum separation, but not a problem that is defined 
for every input size $N$, as is more customary. We know, however, 
that for every such matrix $M$, the tensor products $M^{\ox t}$ 
are also $W(n^t,k^t)$ weighing matrices (for all $t \in \N$). 
We therefore have the  following direct consequence of our results.
\begin{corollary}
Every weighing matrix $M\in \W(n,k)$ defines an infinite family
of $\W(N,K)$ weighing matrix problems, with parameters 
$N=n^t$ and $K=k^t=N^{\log_n k}$ for every $t\in\N$.
By defining $\gamma = 1-\log_n k$
we have, for every suitable $N$, a quantum algorithm with query
complexity $\smfrac{\pi}{4}\sqrt{N^{\gamma}}$ for which there 
is a classical, probabilistic lower bound of 
$(1-\varepsilon)\cdot N^{\gamma}+o(1)$.
\end{corollary}

\begin{example}
Using the $\W(4^t,3^t)$ weighing matrices of Example~\ref{ex:w43},
we have $\gamma=1-\smfrac{1}{2}\log 3 \approx 0.21$, and hence 
a quantum algorithm with query complexity 
$\smfrac{\pi}{4}N^{0.10\dots}$. The corresponding classical 
probabilistic, lower bound of this problem is 
$(1-\varepsilon)\cdot N^{0.21\dots}+o(1)$.
\end{example}

A legitimate objection against the Weighing Matrix Problem is that
it does not seem to be very useful for solving real-life problems. 
In order to obtain more natural problems one can try to look into 
the specific structure that constitutes the weighing matrix
or matrices.  An example of such an approach will be given in 
the next section via Paley's construction of Hadamard 
matrices.  We will see how this leads to the definition of a
problem dealing with quadratic residues of finite fields 
that has a quantum solution that is more efficient than any 
classical protocol.

\section{The Shifted Legendre Sequence Problem}\label{sec:SLSP}
The query task that we will define in this section relies 
on some standard properties of finite fields.
See the appendix of this article for a short but sufficient 
overview of this theory.  Especially important is the
following function which generalizes the Legendre symbol
$(\smfrac{i}{p})$ over $\Z/(p\Z)$ to all finite fields $\F_q$.
(From now on, $p$ will alway denote an odd prime, and $q=p^k$
a power of such a prime.)

\begin{definition}[Legendre symbol over a Finite Field]\label{def:ls}
For every finite field $\F_{q}$, with $q=p^k$ an
odd prime power, the Legendre symbol function 
$\chi:\F_{q}\rightarrow \{-1,0,+1\}$ indicates if 
a number is a quadratic residue or not:
\begin{eqnarray*}
\chi(i)
& := &
\left\{{
\begin{array}{rl}
0 & \mbox{if  $i = 0$}\\
+1 & \mbox{if  $\exists j\neq 0: j^2  = i $}\\
-1 & \mbox{if  $\forall j: j^2 \neq i$.} \\
\end{array}
}\right.
\end{eqnarray*}
\end{definition}

This function is a quadratic, multiplicative character over $\F_q$,
which implies the following result.
\begin{fact}[Near Orthogonality of Shifted Legendre Sequences]\label{fc:no}
For the `inner product' between two Legendre sequences that are shifted
by $s$ and $r\in\F_{q}$ it holds that
\begin{eqnarray*}
\sum_{i \in \F_{q}}{\chi(i+r)\chi(i+s)} & = &
\left\{
{\begin{array}{rl}
q-1 & \mbox{if $s=r$,} \\
-1 & \mbox{if $s\neq r$}.\\
\end{array}}
\right.
\end{eqnarray*}
\end{fact}
\begin{proof}
See the proof of Fact~\ref{fc:inner_chi} in the appendix.
\end{proof}

Raymond Paley used this near orthogonality property 
for the construction of his Hadamard matrices.\cite{P:oom}
Here we will use the same property to describe a problem that,
much the like the weighing matrix problem of the previous section, 
has a clear gap between its quantum and classical query complexity.
In light of Theorem~\ref{thm:qupperb} and Lemma~\ref{lm:clowerb} 
the results of this section are probably not very surprising.
Rather, we wish to give an example of how we can borrow the ideas
behind the construction of combinatorial objects for the design
of new quantum algorithms.  
In this case this is done by stating a problem that uses 
the Legendre symbol over finite fields.
 
\begin{definition}[Shifted Legendre Sequence/SLS Problem]\label{def:slsp}
Assume that we have a black-box for a shifted Legendre function 
$f_s: \F_{q}\rightarrow \{-1,0,+1\}$ that obeys
$f_s(i) := \chi(i+s)$,
with the---for us unknown---shift parameter $s\in \F_{q}$.
The task is to determine (with probability $1-\varepsilon$) 
the value $s$ with a minimum number of calls to the 
function $f$.
\end{definition}

\subsection{Classical Query Complexity of the SLS Problem}
Before describing the quantum algorithm for the Shifted Legendre
Sequence Problem, we will first determine its classical query 
complexity.
The following lower bound is established in a way
similar to the proof of Lemma~\ref{lm:clowerb}.

\begin{lemma}[Classical lower bound SLS Problem]
Assume a classical algorithm that tries to solve the 
shifted Legendre sequence problem over a finite field $\F_{q}$.
To determine the requested value $s$ with a maximum error rate 
$\varepsilon$, requires more than $\log q + \log(\smfrac{1-\varepsilon}{2})$ 
queries to the function $f_s$.
\end{lemma}
\begin{proof}
Consider, like in the proof of Lemma~\ref{lm:clowerb}, a 
decision tree with nodes $v$ and corresponding 
query indices $i_v$.
For every index $i_v$ there is exactly one function
with $f(i_v)=0$. For the tree 
this implies that every node $v$ can only have two proper subtrees 
(corresponding to the answers $f(i)=+1$ and $-1$) and one 
deciding leaf (the case $f_{(-i)}(i)=0$).
Hence, a decision tree of depth $d$ can distinguish no more than 
$2^{d+1}-1$ different functions.
In order to be able to differentiate between $(1-\varepsilon)q$ functions, 
we thus need a depth $d$ of at least $\log((1-\varepsilon)q+1)-1$.
\end{proof}

This lower bound of $\log q$ queries is also optimal as 
is shown by the following result.

\begin{lemma}[Classical upper bound SLS Problem]
There exists a deterministic, classical protocol that 
solves the Shifted Legendre Sequence Problem with 
$O(\log q)$ queries to the black-box function $f_s$.
\end{lemma}
\begin{proof}
Let $S\subseteq \F_q$ be the set of possible values 
of $s$ at a given moment during the execution of the protocol.
(Thus, initially we  have $S=\F_q$, and we want to end 
with the unique answer determined by $|S|=1$.)  
Below we will show that if $S$ has at least $4$ 
elements, then there always exists an index $i$ 
such that all three possible answers to the query 
``$f(i)$?'' lead to a reduced the set of options 
$S'$ with $|S'| < \smfrac{3}{4}|S|$.
By repeating this procedure no more than 
$\log q/\log(\smfrac{4}{3})$ times, 
we can reduce the initial set $S=\F_q$ to four 
possibilities, which can then be checked with three 
additional queries.  

What follows is the existence proof of such an index
for every possible subset $S\subseteq \F_q$.
Given a set $S$ and an index $i$, we have 
a partition of $S$ in three subsets according to
the answer $\chi(s+i)$ to the query ``$f(i)$?'' 
\begin{eqnarray*}
S_i^+ & := & \{j |j\in S\textrm{ and }\chi(s+j)=+1 \}, \\
S_i^- & := & \{j |j\in S\textrm{ and }\chi(s+j)=-1 \}, \\
S_i^0 & := & \{j |j\in S\textrm{ and }\chi(s+j)=0 \}.
\end{eqnarray*}
Note that, depending on whether $-s$ is an element of $S$ 
or not, $S_i^0$ is either $\{-s\}$ or the empty set.
Clearly, one of these three sets will be the 
 reduced set $S'$ mentioned in the first part of the proof.

Define the Legendre matrix $L\in\{-1,0,+1\}^{q\times q}$ 
by $L_{ij} := \chi(i+j)$, and let $z_S$ be the characteristic 
vector $\in \{0,1\}^q$ of the subset $S\subseteq\F_q$. 
The product of $L$ and $z_S$ yields a new vector $w_s$ with
the following property for its $i$-th entry:
\begin{equation*}
(w_S)_i \quad = \quad \sum_{j\in\F_q}{\chi(i+j)(z_S)_j}\quad
= \quad  \sum_{j\in S}{\chi(i+j)}\quad  =\quad  |S_i^+|-|S_i^-|.
\end{equation*}
By the near orthogonality property of the Legendre sequence 
(Fact~\ref{fc:no}) we know that for the matrix $L$ we 
have $L^T\cdot L = q\I_q - \J_q$, where $\J_q$ is the `all ones'
matrix of dimension $q\times q$. 
The inner product $z_S^T z_S$ is the Hamming weight of $z_S$
and hence equals the size $|S|$ of the set $S$.
This implies for the inner product of $Lz_S$ with itself:
$(z_S^T L^T)(L z_S) = z_S^T(q\I_q - \J_q)z_S = q |S| - |S|^2$.
By the previous equation for $w_S$, we thus see that
\begin{equation*}
q|S|-|S|^2 \quad = \quad w_S^Tw_S \quad = \quad
\sum_{i\in\F_q}{\left(|S^+_i|-|S^-_i|\right)^2}.
\end{equation*}
This proves that there exist at least one index $i$ 
for which $(|S_i^+|-|S_i^-|)^2 \leq \smfrac{1}{q}(q|S|-|S|^2)$, 
and hence $-\sqrt{|S|} < |S_i^+|-|S_i^-| < \sqrt{|S|}$.
In combination with the general bound 
$|S_i^+|+|S_i^-|\leq |S|$, this gives that for this
$i$ both $|S^+_i|$ and $|S^-_i|$ are less than  
$\smfrac{1}{2}|S|+\smfrac{1}{2}\sqrt{|S|}$.
For $|S|\geq 4$ this proves that indeed all three 
$S_i^+$, $S_i^-$ and $S^0_i$ have size
less than $\smfrac{3}{4}|S|$.
\end{proof}

\subsection{Quantum Query Complexity of the SLS Problem}
The above classical upper bound for the Shifted Legendre Sequence
Problem relied on the near orthogonality property of the Legendre 
sequence.  The same is done by the quantum protocol for the SLS
Problem, but in a more efficient way: only $2$ quantum
queries are required.
\begin{theorem}[Two query quantum protocol for the SLS Problem]
\label{thm:qslsp}
For any finite field $\F_{q}$, the Shifted Legendre Sequence Problem 
of Definition~\ref{def:slsp} can be solved exactly with two quantum 
queries to the black-box  function $f_s$. 
\end{theorem}
\begin{proof}
We exhibit the quantum algorithm in detail.
We start with the superposition
\begin{equation*}
\frac{1}{\sqrt{q + 1}}\left({
\sum_{i \in \F_{q}}{\ket{i}\ket{0}}
\quad  + \quad \ket{\textrm{dummy}}\ket{1}
}\right).
\end{equation*}
(The reason for the ``dummy'' part of state that we use will
be clear later in the analysis.)
The first oracle call is used to calculate the different
$f_s(i)=\chi(i+s)$ values for the non-dummy states, giving
a superposition of states $\ket{i,\chi(i+s)}$.
At this point, we measure the rightmost register to see if it
contains the value ``zero''.
If this is indeed the case (probability $\smfrac{1}{q+1}$), the
state has collapsed to $\ket{-s}\ket{0}$ which directly gives us
the desired answer $s$.
Otherwise, we continue with the now reduced state
\begin{equation*}
\frac{1}{\sqrt{q}}\left({
\sum_{i \in \F_{q}\!\setminus\!\{-s\}}{\ket{i}\ket{\chi(i+s)}}
\quad  + \quad \ket{\textrm{dummy}}\ket{1}
}\right),
\end{equation*}
on which we apply a conditional phase change (depending on the
$\chi$ values in the rightmost register).
We finish the computation by `erasing' this rightmost register
with a second call to $f_s$. (For the dummy part, we just reset
the value to ``zero''.)
This gives us the final state $\psi$, depending on $s$, of the form
\begin{eqnarray*}
\ket{\psi_s}\ket{0} & := &
\frac{1}{\sqrt{q}}\left({
\sum_{i \in \F_{q}}{\chi(i+s)\ket{i}}
\quad  + \quad \ket{\textrm{dummy}}
}\right)\ket{0}.
\end{eqnarray*}

What is left to show is that $\{\ket{\psi_s}|s\in\F_{q}\}$ forms a
 set of orthogonal vectors.
Fact~\ref{fc:no} tells us that for the inner product
between two states $\psi_s$ and $\psi_r$ it holds that
$\braket{\psi_r}{\psi_s} = 1$ if $s= r$, and
$\braket{\psi_r}{\psi_s} = 0$ if $s\neq r$.
 In other words, the states $\psi_s$ for $s\in \F_{q}$ 
are mutually orthogonal.
Hence, by measuring the final state in the $\psi$-basis, we can  
determine without error the shift factor $s\in\F_{q}$ after only
two oracle calls to the function $f_s$.
\end{proof}
More recently, Peter H{\o}yer has shown the existence of a one 
query protocol for the same problem [private communication].

\subsection{Query versus Time Complexity Issues}
The above algorithm only reduces the \emph{query complexity} to
$f_s$. The \emph{time complexity} of the protocol is another matter,
as we did not explain how to perform the final measurement along the
$\psi$ axes in a time-efficient way. 
This question, whether there
exists a tractable implementation of the unitary mapping
\begin{eqnarray*}
\ket{s}
& \longleftrightarrow &
\frac{1}{\sqrt{q}}\left({
\sum_{x \in \F_{q}}{\chi(x+s)\ket{x}}
\quad + \quad \ket{\textrm{dummy}}
}\right),
\end{eqnarray*}
is discussed and solved in an independent article\cite{DamHallgren}.

\section{Conclusion}
We have established a connection between the construction
of weighing matrices in combinatorics, and the design of 
new quantum algorithms.
  It was shown how every weighing 
matrix leads to a query problem that has a more efficient
quantum solution than is possible classically.
The earlier known results of Bernstein \& Vazirani\cite{BV:qct}
and Grover\cite{G:afqmafds} were shown to be specific instances of 
this more general problem.

Starting from Paley's construction of Hadamard matrices\cite{H:rduqrad}, 
we used the structure of quadratic residues over finite
fields to give a more explicit example of a Weighing Matrix Problem.
This led to the definition of the Shifted Legendre Sequence Problem,
which has a constant quantum query complexity, compared to 
a logarithmic classical query complexity.

Although the results in this article only concern the
\emph{query complexity} of black-box problems, it 
should be viewed as the first step towards the construction
of quantum protocols that are also time-efficient and
that deal with more realistic problems. 
Constructions of Hadamard matrices that are especially 
interesting in this context are, for example, the complex Hadamard 
matrices of Turyn\cite{T:chm} and the Hadamard matrices of the dihedral 
group type\cite{K:hmadg,SY:afohmodgt}.
Also, the time-efficiency of the quantum 
solution of (a generalization of) the Shifted Legendre 
Sequence Problem was proven more recently in \cite{DamHallgren}.

\section*{Acknowledgments}
Harry Buhrman, Peter H{\o}yer, Paul Vit\'anyi and Ronald de Wolf are 
thanked for several useful discussions. 
Mike Mosca and Alain Tapp are especially acknowledged for their
valid criticism of an earlier version of this article,
which was made possible by the hospitality of the 
CACR at the University of Waterloo.

This work has been supported by the Institute for Logic, Language 
and Computation in Amsterdam, the EU fifth framework project 
QAIP IST-1999-11234, and the TALENT grant S 62-552 of the
Netherlands Organization for Scientific research (NWO).

\appendix
\section{Quadratic Residues of Finite Fields}\label{sec:ff}
This appendix describes some standard results about quadratic residues
and Legendre symbols over finite fields.
For more background information one can look up  
 references like \cite{C:acicant} or \cite{IR:acitmnt}.

\subsection{Finite Field Factoids}
Let $q=p^k$ denotes a power of an odd prime $p$.
There always exists a generator $\zeta$ for the multiplicative 
group $\F_{q}^{\star} = \F_{q}\!\setminus\!\{0\}$.
This means that the sequence $\zeta, \zeta^2, \zeta^3,\dots$
will generate all non-zero elements of $\F_{q}$.
As this is a set of size $q-1$, it follows that
$\zeta^{q} = \zeta$, and hence $\zeta^{(q-1)} = 1$.
Hence we have the equivalence relation
\begin{eqnarray*}
\zeta^{i} = \zeta^{j}
& \textrm{ if and only if }&
i = j \bmod{(q-1)}
\end{eqnarray*}
for every integer $i$ and $j$.

We now turn our attention to the definition of the \emph{generalized 
Legendre symbol} $\chi:\F_q\rightarrow\{-1,0,+1\}$ which is defined by:
\begin{eqnarray*}
\chi(i)
& := &
\left\{{
\begin{array}{rl}
0 & \textrm{ if  $i = 0$}\\
+1 & \textrm{ if  $\exists j\neq 0: j^2  = i $}\\
-1 & \textrm{ if  $\forall j: j^2 \neq i$.} \\
\end{array}
}\right.
\end{eqnarray*}
By the above mentioned equivalence relation, the quadratic expression
$(\zeta^j)^2 = \zeta^{2j}=\zeta^i$ is correct if and only if
$2j = i \bmod{q-1}$. As $p$ is odd, $q-1$ will be even,
and hence there can only exists a $j$ with $(\zeta^j)^2 = \zeta^i$
when $i$ is even. Obviously, if $i$ is even, then
$\zeta^j$ with $j=\smfrac{i}{2}$ gives a solution to our quadratic
equation. This proves that $50\%$ of the elements
of $\F_{q}^{\star}$ are a quadratic residue with $\chi(x)=+1$,
while the other half has $\chi(x)=-1$.
In short: $\chi(\zeta^i) = (-1)^i$, and hence for the total
sum of the function values: $\sum_x{\chi(x)}=0$.

\subsection{Multiplicative Characters over Finite Fields}
The rule $\chi(\zeta^i)\cdot\chi(\zeta^j) = \chi(\zeta^{i+j})$,
in combination with $\chi(0)=0$, shows that the Legendre symbol 
$\chi$ is a \emph{multiplicative character}
with $\chi(x)\cdot\chi(y)=\chi(xy)$ for all $x,y \in \F_{q}$.
\begin{definition}[Multiplicative characters over finite fields]
The function $\chi:\F_{q}\rightarrow \C$ is a \emph{multiplicative
character} if and only if $\chi(xy) = \chi(x)\chi(y)$ for 
all $x,y \in \F_{q}$. 
The constant function $\chi(x)=1$ is called the trivial character. 
(We do not consider the other trivial function $\chi(x)=0$.)
\end{definition}
See \cite{C:acicant,IR:acitmnt} for the usage of multiplicative
characters in number theory.  Some of the properties that we 
will use in this article are: $\chi(1)=1$,
if $\chi$ is nontrivial, then $\chi(0)=0$,
the inverse of nonzero $x$ obeys $\chi(x^{-1}) = {\chi(x)}^{-1} = \overline{\chi(x)}$,
and $\sum_x \chi(x) = 0$ for nontrivial $\chi$.

With these properties we can prove the following fact that we
use in the article.
\begin{fact}[Near orthogonality of shifted characters]\label{fc:inner_chi}
Consider a nontrivial  character $\chi:\F_{q}\rightarrow\C$.
For the `complex inner product' between two $\chi$-sequences 
that are shifted by $s$ and $r\in\F_{q}$ it holds that
\begin{eqnarray*}
\sum_{x \in \F_{q}}{\overline{\chi(x+r)}\chi(x+s)} & = &
\left\{
{\begin{array}{rl}
q-1 & \mbox{if $s=r$,} \\
-1 & \mbox{if $s\neq r$}.\\
\end{array}}
\right.
\end{eqnarray*}
\end{fact}
\begin{proof}
(For the quadratic character, these are simple instances of 
so-called \emph{Jacobsthal sums}\cite{J:udddpdf4aszq}; 
see for example Section 6.1 in \cite{BEW:gajs}.)
Rewrite
\begin{eqnarray*}
\sum_{x \in \F_{q}}{\overline{\chi(x+r)}\chi(x+s)} &= &
\sum_{x \in \F_{q}}{\overline{\chi(x)}\chi(x+\Delta)}
\end{eqnarray*}
with $\Delta := s-r$. If $s=r$ this sum equals $q-1$. Otherwise,
we can use the fact that
$\overline{\chi(x)}\chi(x+\Delta) = 
\chi(1+x^{-1}\Delta) = 
\chi(\Delta)\chi(\Delta^{-1}+x^{-1})$
(for $x\neq 0$) to reach
\begin{eqnarray*}
\sum_{x \in \F_{q}}{\overline{\chi(x)}\chi(x+\Delta)}& =& 
\chi(\Delta)\sum_{x \in {\F_{q}}^{\star}}{(\Delta^{-1}+x^{-1})}.
\end{eqnarray*}
Earlier we noticed that $\sum_x{\chi(x)}=0$,
and therefore in the above summation (where the value $x=0$ is omitted)
we have $\sum_{x}{\chi(x^{-1}+\Delta^{-1})} = -\chi(\Delta^{-1})$.

This confirms that indeed
\begin{eqnarray*}
\chi(\Delta)\sum_{x \in {\F_{q}}^{\star}}{\chi(x^{-1}+\Delta^{-1})} &=& -1,
\end{eqnarray*}
which finishes the proof.
\end{proof}
Note that the above property should not be confused with the orthogonality 
relation 
\begin{eqnarray*}
\sum_{x\in\F_q} {\overline{\chi(x)}\chi'(x)} & = & 
\left\{
\begin{array}{rl}
q-1 & \mbox{if $\chi=\chi'$} \\
0 & \mbox{if $\chi\neq\chi'$} \\
\end{array}
\right.
\end{eqnarray*}
between two (possibly different) characters $\chi$ and $\chi'$.
\end{document}